\documentclass[preprint,showpacs,preprintnumbers,amsmath,amssymb]{revtex4}

\usepackage{amsmath}
\usepackage{amssymb}
\usepackage{graphicx} 
\usepackage{color}
\usepackage{dcolumn}
\usepackage{bm}
\usepackage[latin1]{inputenc}

\newcommand{\bx}{\mathbf{x}}
\newcommand{\R}{\mathbb{R}}

\newcommand{\C}{\mathbb{C}}

\newcommand{\gH}{\mathfrak{h}}
\newcommand{\norm}[1]{ | \! | #1 | \! | }
\newcommand{\tr}{{\rm tr}}

\newcommand{\cS}{\mathcal{S}}
\newcommand{\cP}{\mathcal{P}}

\newcommand{\cL}{\mathcal{L}}
\newcommand{\cC}{\mathcal{C}}
\newcommand\pscal[1]{{\ensuremath{\langle #1 \rangle}}}
\newcommand{\splus}{|\!\uparrow\rangle}
\newcommand{\smoins}{|\!\downarrow\rangle}

\newtheorem{algo}{Algorithm}

\begin{document}

\preprint{APS/}

\title{The Electronic Ground State Energy Problem: \\
a New Reduced Density Matrix Approach}

\author{Eric Canc\`es}
\email{cances@cermics.enpc.fr}
\altaffiliation[Also at ]{INRIA, Rocquencourt.}
\author{Gabriel Stoltz}
 \email{stoltz@cermics.enpc.fr}
\altaffiliation[Also at ]{CEA/DAM, Bruy\`eres-le-Châtel.}
 \affiliation{%
CERMICS, Ecole des Ponts, ParisTech, France
}%
\author{Mathieu Lewin}%
 \email{lewin@math.cnrs.fr}
\affiliation{%
CNRS \& Universit\'e de Cergy-Pontoise, France
}%

\date{\today}

\begin{abstract}
We present here a formulation of the electronic ground-state
energy in terms of the second order reduced density matrix, using a
duality argument.   
It is shown that the computation of the ground-state energy reduces to
the search of the projection of some two-electron reduced Hamiltonian on
the dual cone
of $N$-representability conditions. Some numerical results
validate the approach, both for equilibrium geometries and for the
dissociation curve of N$_2$. 
\end{abstract}

\pacs{
}
\maketitle


\section{Introduction}

As early as in 1951, it was noticed by Coleman that the electronic $N$-body
ground-state energy could be obtained by minimizing over the set of
$N$-representable two-body reduced density matrices (2-RDM), 
and Mayer definitely opened the field in 1955 with his pioneering article~\cite{mayer55}.
At a conference in 1959, Coulson then proposed to completely eliminate wavefunctions from
Quantum Chemisty, since all the electronic ground-state properties of molecular systems can be computed from the 2-RDM~\cite{coulson60,lowdin55,mayer55}. Unfortunately, the set of 
$N$-representable 2-RDM is not known explicitly. Some
mathematical characterizations were provided
\cite{Kummer,Coleman-Yukalov,Coleman02} but they could not be used to
derive a numerical method with a complexity of a lower order than the
usual $N$-body problem. Analytical approaches for model systems (see e.g.~\cite{Percus78}) were also proposed in order to precise the accuracy of the $N$-representability conditions in specific cases. In practice, only \emph{approximate} RDM minimization problems, in which only a few necessary $N$-representability conditions are imposed (see the geometric constraints of~\cite{yamada60}, or the so-called P,Q,G conditions~\cite{Coleman63,GP64}), can be considered. The first numerical studies relying on this strategy gave encouraging results~\cite{GMR75}. 

Recently a new interest in the Reduced Density Matrix (RDM) approach
arose. Impressive numerical results have been obtained by
two different strategies issued from semidefinite 
programming: primal-dual interior point methods~\cite{NNEFNF01,mazziotti02,ZBFOP,FBNOPYZ} on the one hand, augmented Lagrangian formulations using matrix factorizations
of the 2-RDM~\cite{mazziottiPRL,mazziotti04,mazziotti05} on the other hand. These results use a small number of known \emph{necessary conditions} of $N$-representability.
Yet, the so-obtained ground-state energies are as accurate as
the ones obtained with coupled-cluster methods,  see
e.g. \cite{mazziottiPRL,mazziotti04}. In addition, these energies provide
lower bounds of the Full CI energies, whereas the variational
post Hartree-Fock methods, such as CI or MCSCF, all provide upper
bounds.

Although the current implementations of variational 2-RDM algorithms are limited to the simulation of small molecules in small basis sets, we believe that improvements of the algorithms and increase of computational power will make it possible to simulate larger molecules and to use larger basis sets in a near future. This will allow in particular to assess the convergence of the RDM approach with respect to the size of the basis set, for a given molecular system.

Since the RDM method is a linear minimization problem over a
convex set of complicated structure, it is natural to use the concept of
duality to 
mathematically characterize and numerically compute the minimum. Duality
is an underlying issue in all the RDM studies
\cite{Kummer,GP64,Erdahl,Erdahl2,Coleman-Yukalov,Coleman02}, but
surprisingly, the specific form of the dual formulation of the RDM
problem has not yet been used to derive an efficient algorithm. The
current methods (see,
e.g. \cite{NNEFNF01,ZBFOP,FBNOPYZ,mazziottiPRL,mazziotti04}) all use general
duality considerations in their algorithms, but none of them solves
directly (and only) the dual RDM problem. The purpose of the present
article is to present such an approach. As will be shown below, the
associated dual optimization problem 
boils down to the search of the zero of a one-dimensional convex
function. 

The paper is organized as follows. After
setting the problem in section~\ref{section_notation}, we derive the RDM and the approximate RDM dual problems by standard Lagrangian methods in section~\ref{section_dual}. Then, in section~\ref{section_algo}, we propose a new algorithm which aims at solving directly the dual problem. Section~\ref{section_num} eventually presents some numerical results demonstrating that this new method is an interesting and efficient alternative to the existing methods. 


\section{Notation}
\label{section_notation}

Let us consider a finite-dimensional space $\gH:={\rm span}(\chi_i,\
i=1,...,r)$ 
where $(\chi_i)_{i\geq1}$ is a Hilbert basis of the
one-body space $L^2(\R^3\times \{\splus, \smoins\},\C)$. Most of our
analysis is also valid in infinite dimension but  for the sake of
simplicity, we restrict
to the finite-dimensional case. The electronic
Hamiltonian $H_N$ acts on the $N$-body fermionic space
$\bigwedge_{n=1}^N\gH$ of 
antisymmetric $N$-body wavefunctions $\Psi(x_1,...,x_N)$ and is formally
defined as
$$
H_N=\sum_{i=1}^Nh_{x_i}+\sum_{1\leq i<j\leq N}\frac{1}{|\bx_i-\bx_j|}
$$
where $h=-\Delta/2+V$ and $V$ is the external Coulomb potential
generated by the nuclei. In the whole paper, we denote by
$x=(\bx,\sigma)$ the vector containing both the space variable
$\bx\in\R^3$ and the spin variable $\sigma\in\{\splus, \smoins\}$. For
any vector space $X$, we denote by $\cS(X)$ the space of self-adjoint
matrices acting on $X$, and by $\cP(X)\subset\cS(X)$ the cone of
positive semi-definite matrices. We also use the simplified notation
$\cP_N:=\cP\left(\bigwedge_1^N\gH\right)$ and
$\cS_N:=\cS\left(\bigwedge_1^N\gH\right)$. The ground-state energy then reads
\begin{equation}
E=\inf_{\substack{\Psi\in\bigwedge_{n=1}^N\gH,\\ \norm{\Psi}=1}}\pscal{\Psi,H_N\Psi}=\inf_{\substack{\Upsilon\in\cP_N,\\
    \tr(\Upsilon)=1}}\tr(H_N{\Upsilon}). 
\label{def_ground_state}
\end{equation} 
The second equality holds true for the infimum of the energy over
the set of 
mixed states coincides with the infimum of the energy over the
set of pure states. In mathematical words, the minimum of a linear function over a convex set
is attained on an extremal point of the convex set.

The 2-RDM $\Gamma$ associated with an $N$-body density matrix $\Upsilon\in\cP_N$ is defined by means of Kummer's contraction operator $L_N^2$ as \cite{Kummer,Coleman-Yukalov}
\begin{equation}
\Gamma_{i_1,i_2}^{j_1,j_2}=L_N^2(\Upsilon)_{i_1,i_2}^{j_1,j_2}=N(N-1)\sum_{k_3,...,k_N=1}^r\Upsilon_{i_1i_2k_3...k_N}^{j_1j_2k_3...k_N}.
\end{equation}
Then, the \emph{cone $\cC_N$ of $N$-representable two-body density matrices} is by definition the image by $L_N^2$ of the cone $\cP_N$ of $N$-body density matrices:
$$\cC_N=L_N^2(\cP_N)\subset\cS_2.$$
Of course the 2-RDMs of physical interest are the elements $\Gamma\in\cC_N$ which arise from a normalized $N$-body density matrix $\Upsilon$, i.e. which additionally satisfy $\tr(\Gamma)=N(N-1)$.

Since the Hamiltonian $H_N$ only contains two-body interactions, the energy of the system can be expressed in terms of the
two-body density matrix $\Gamma$ only (see, e.g. \cite{Coleman-Yukalov,mazziotti02}): 
\begin{equation}
\label{inf}
\fbox{$\displaystyle E = \inf_{\substack{\Gamma\in \cC_N,\\ \tr(\Gamma)=N(N-1)}}\tr(K_N \Gamma)$}
\end{equation}
where we have introduced
$$K_N=\frac{h_{x_1}+h_{x_2}}{2(N-1)}+\frac{1}{2|\bx_1-\bx_2|}.$$ 
Formula \eqref{inf} is an obvious consequence of the identity $H_N=(L^2_N)^*K_N$ where $(L^2_N)^*$ is the adjoint of $L^2_N$ sometimes also called a \emph{lifting operator}.
Notice that we did not impose any constraint on the spin state in~\eqref{inf}, but such
constraints can be easily taken into account. 

\section{Dual Formulation of the RDM Minimization Problem}
\label{section_dual}

We now present the dual formulation of the minimization \eqref{inf}. We
recall that the polar cone $\cC^*$ of a cone $\cC$ in any Hermitian space
is defined as $\cC^*=\{x\ |\ \forall y\in\cC, \pscal{x,y}\geq0\}$, where
$ \pscal{\cdot,\cdot}$ denotes the considered scalar product.
The dual method then consists in formulating \eqref{inf} in terms of $(\cC_N)^*$ instead of $\cC_N$:
\begin{equation}
\label{dual}
\fbox{$\displaystyle E=N(N-1)\sup\{\mu\ |\ K_N-\mu\in(\cC_N)^*\}.$}
\end{equation}
We therefore obtain an optimization problem in dimension 1 over $\mu\in\R$ which is the variable dual to the constraint $\tr(\Gamma)=N(N-1)$.
 Of course characterizing the polar cone $(\cC_N)^*$ is as difficult as
 characterizing $\cC_N$, this issue is called the \emph{$N$-representability problem}. Indeed $\cC_N=(\cC_N)^{**}$. Even if the dual
 formulation \eqref{dual} does not simplify the theoretical $N$-representability
 problem, it turns out to be more convenient for numerical purposes, as
 will be shown below. 

Formula~\eqref{dual} can be easily derived from \eqref{inf}. Introducing the Lagrangian
$$\cL(\Gamma,B,\mu)=\tr(K_N\Gamma)-\tr(B\Gamma)-\mu\{\tr(\Gamma)-N(N-1)\},$$
it follows
\begin{equation}
E=\inf_{\Gamma\in\cS_2}\sup_{B\in(\cC_N)^*,\ \mu\in\R}\cL(\Gamma,B,\mu).
\label{Lagrangian_form}
\end{equation}
 It then suffices to exchange the $\inf$ and the $\sup$ in \eqref{Lagrangian_form} to obtain \eqref{dual}. Indeed, it is a general fact that for any cone $\cC$ in a finite-dimensional space
\begin{equation}
\inf_{x\in\cC,\ \pscal{b,x}=1}\pscal{a,x}=\sup\{\mu\ |\ a-b\mu\in\cC^*\}.
\label{general_dual}
\end{equation}
Note that this property has been already used in the RDM setting by Erdahl
\cite{Erdahl2}. We shall use it again below.

Since both $(\cC_N)^*$ and $\cC_N$ are unknown and difficult to characterize, it is necessary to
approximate \eqref{dual} by a variational problem that can be carried
out numerically. To this end, some necessary conditions for
$N$-representability are selected. We consider in this paper $L$
conditions of the following general form 
\begin{equation}
\forall \ell=1...L,\quad \cL_\ell(\Gamma)\geq0
\label{conditions}
\end{equation}
where for any $\ell$, $\cL_\ell:\cS_2\to \cS(X_\ell)$ is a linear map and $X_\ell$
is some vector space. For instance, the so-called ${\rm P}$-condition
$\cL_1(\Gamma)=\Gamma 
\geq 0$ (with $X_1=\gH\wedge\gH$) originates from the Kummer
operator preserving positivity, and will always be considered. Other classical necessary conditions of
$N$-representability will be introduced
below. Imposing only the necessary conditions \eqref{conditions} means
that $\cC_N$ is replaced by the approximate cone $\cC_{\rm
  app}\supset\cC_N$ defined as 
\[
\cC_{\rm app}:= \{\Gamma\in \cS_2\ |\ \forall \ell=1...L,\ \cL_\ell(\Gamma)\geq0\}.
\]
Its polar cone can easily be shown to be
\begin{equation}
(\cC_{\rm app})^*:= \left\{\sum_{\ell=1}^L(\cL_\ell)^*B_\ell\ |\ B_\ell\in\cS(X_\ell),\ B_\ell\geq0\right\},
\label{app_dual_cone}
\end{equation}
and the associated approximate energy is then, in view of~\eqref{general_dual},
\begin{eqnarray}
E_{\rm app} & = & \inf_{\substack{\Gamma\in \cC_{\rm app},\\ \tr(\Gamma)=N(N-1)}}\tr(K_N \Gamma)\\
 & = & N(N-1)\sup\{\mu\ |\ K_N-\mu\in(\cC_{\rm app})^*\}\label{app_dual}.
\end{eqnarray}
Let us emphasize that, since $\cC_{\rm app}\supset\cC_N$, the energy
$E_{\rm app}$ is a \emph{lower bound} to the full CI energy in the
chosen basis, $E_{\rm app}\leq E$. We present below an algorithm for
solving problem~\eqref{app_dual}. 
Notice that we obtain only the ground-state energy (and not the ground state density matrix), but, resorting to
first order perturbation theory, any observable including at most two-body interaction terms can be obtained by a finite difference of energies.

Some well-known necessary conditions of the form \eqref{conditions} are
the P, Q, G conditions~\cite{GP64,Coleman-Yukalov}. Additional necessary
conditions can be considered, such as Erdahl's T$_1$ and T$_2$ conditions~\cite{Erdahl,ZBFOP,FBNOPYZ}. The P, Q and G conditions correspond to the following linear operators in \eqref{conditions}: 
$$\cL_1(\Gamma)=\Gamma,$$
\[
\left [ {\cal L}_Q(\Gamma) \right ]_{i_1,i_2}^{j_1,j_2} =  \Gamma_{i_1,i_2}^{j_1,j_2} - \delta_{i_1}^{j_1} \gamma_{i_2}^{j_2} - \delta_{i_2}^{j_2} \gamma_{i_1}^{j_1} + \delta_{i_1}^{j_2} 
\gamma_{i_2}^{j_1} + \delta_{i_2}^{j_1} \gamma_{i_1}^{j_2} 
+ (\delta_{i_1}^{j_1} \delta_{i_2}^{j_2} - \delta_{i_1}^{j_2} \delta_{i_2}^{j_1}) \frac{\tr(\Gamma)}{N(N-1)},
\]
\[
\left [ {\cal L}_G(\Gamma) \right ]_{i_1,i_2}^{j_1,j_2} = -\Gamma_{i_1,j_2}^{j_1,i_2} + \delta_{i_1}^{j_1} \gamma_{i_2}^{j_2},
\]
where $\gamma_i^j = \frac{1}{N-1}\sum_{k=1}^{r} \Gamma_{i,k}^{j,k}$ is
the one-body RDM associated with the two-body RDM $\Gamma$. 
Expressions for the adjoint operators $(\cL_Q)^*$ and $(\cL_G)^*$ were presented in~\cite{mazziotti02} for example.
Notice that
for any $\Gamma\in\cS_2$, $\cL_P(\Gamma)$ and $\cL_Q(\Gamma)$ also are
antisymmetric, whereas $\cL_G(\Gamma)$ is not. Therefore,
$X_P=X_Q=\gH\wedge\gH$ and $X_G=\gH\otimes\gH$ in the above general
formalism. 
Notice also that Erdahl's three-index conditions T$_1$, T$_2$
require $X_{T_i}=\gH\otimes\gH\otimes\gH$. 

Our numerical tests were performed using the P, Q, G conditions but our
algorithm for solving \eqref{app_dual} is valid for any set of necessary
conditions of the form \eqref{conditions}.

\section{Algorithm for solving the dual problem}
\label{section_algo}

\begin{figure}
\begin{center}
\includegraphics[width=16cm,height=10cm]{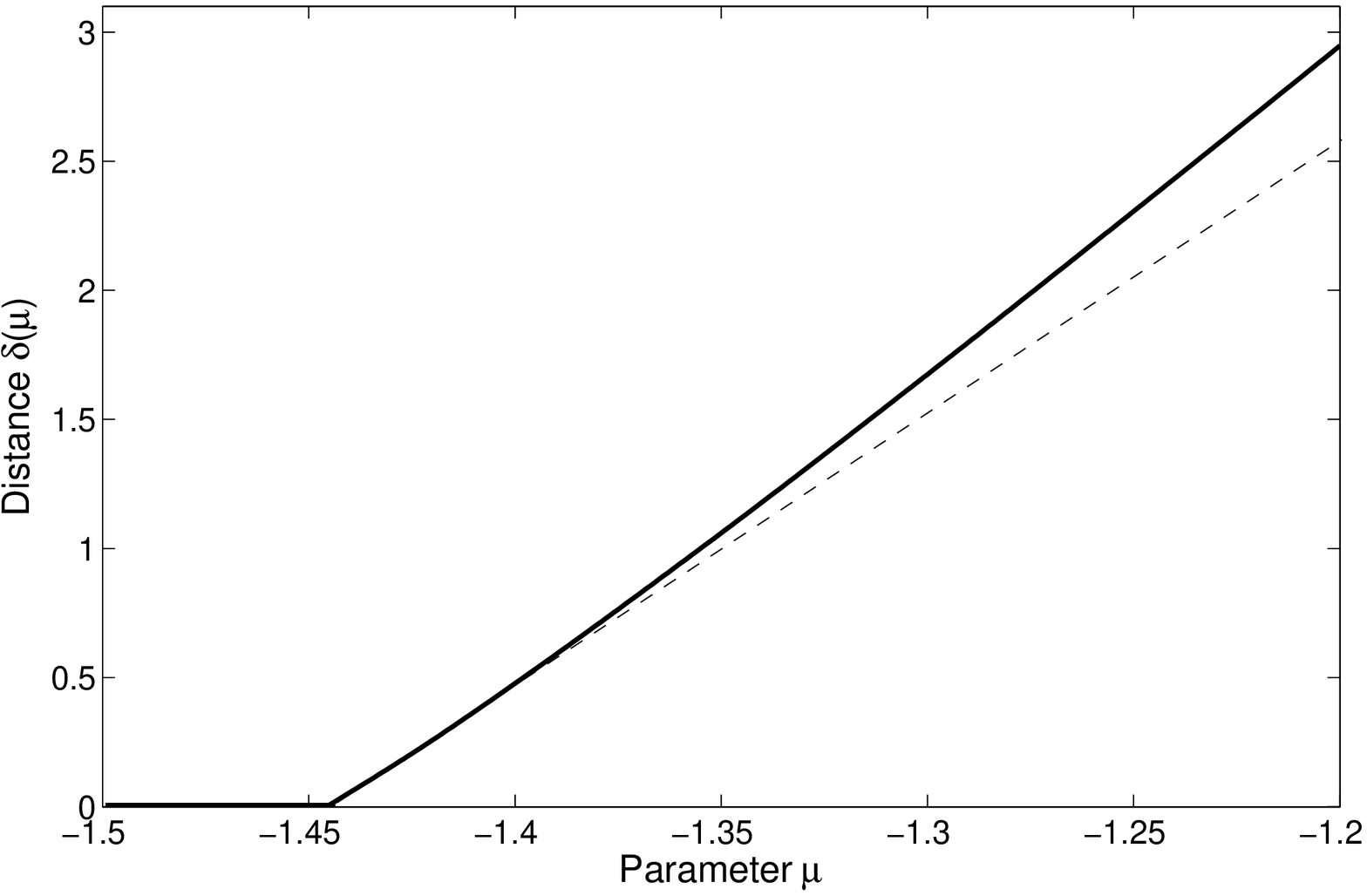}\hfill
\end{center}\vspace{1cm}\begin{center}
\includegraphics[width=16cm,height=10cm]{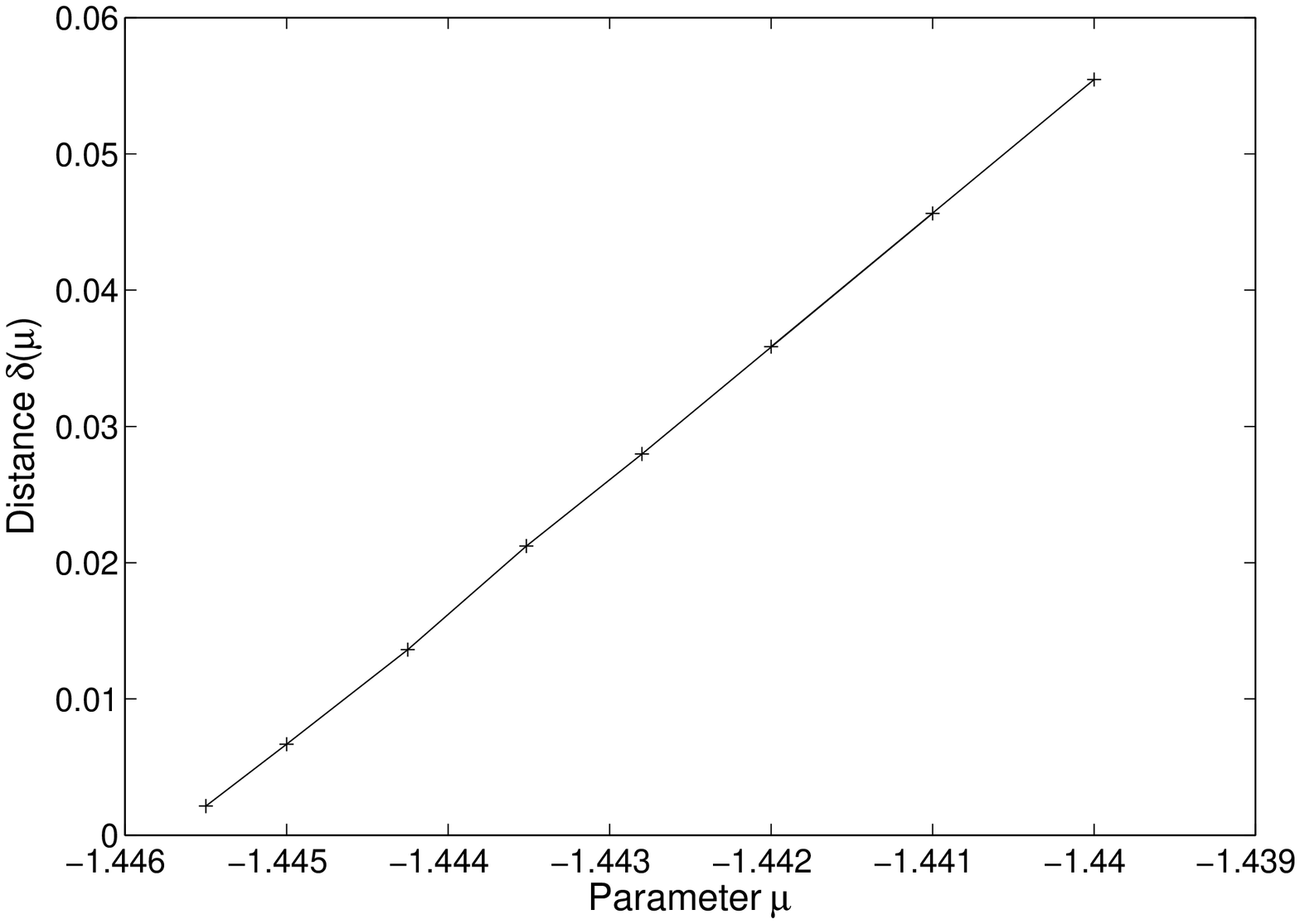}\hfill
\end{center}
\vspace{1.8cm}
\caption{ \label{courbe_N2} Cancès et al., Journal of Chemical Physics.} 
\end{figure}

Let us introduce the distance to the dual cone $(\cC_{\rm app})^*$
$$\delta(\mu)={\rm dist}\left(K_N-\mu,(\cC_{\rm app})^*\right).$$
Denoting $\mu^*_{\rm app}=E_{\rm app}/(N(N-1))$, the function $\delta$ satisfies the following properties:
\begin{description}
\item[$(i)$] $\delta\equiv0$ on $(-\infty,\mu^*_{\rm app}]$ and is increasing on $[\mu^*_{\rm app},\infty)$;
\item[$(ii)$] $\delta$ is convex on $\R$;
\item[$(iii)$] $\delta^2$ is continuously differentiable on $\R$, thus $\delta$ is continuously differentiable on $\R\setminus\{\mu^*_{\rm app}\}$ and
\begin{equation}
\forall \mu>\mu^*_{\rm app},\quad {\delta}'(\mu) =- \frac{\tr(K_N-\mu-A_{\mu})}{\norm{K_N-\mu-A_{\mu}}}
\label{derivee}
\end{equation}
where $A_\mu$ denotes the projection of $K_N-\mu$ onto the polar cone $(\cC_{\rm app})^*$.
\end{description}
Proofs for $(ii)-(iii)$ can be found in \cite{Moreau}. To prove $(i)$, one notices that when $\mu\leq\mu^*_{\rm app}$, $K_N-\mu=K_N-\mu^*+(\mu^*-\mu)$ belongs to $(\cC_{\rm app})^*$ since $\mu^*-\mu\in\cP_2\subset (\cC_{\rm app})^*$.
To illustrate the above properties, we provide a plot of $\delta(\mu)$ for N$_2$ in a STO-6G basis set, see Figure~\ref{courbe_N2}.

\bigskip

{\bf Figure I}

\bigskip

In order to compute $\mu^*_{\rm app}$, we use a Newton-like scheme that 
strongly exploits the above mentioned properties in a natural way:
starting from an initial energy above $\mu^*_{\rm app}$ (such as  the
Hartree-Fock energy for instance) and using the convexity of the
function $\delta$, the Newton algorithm ensures that the energy $\mu$
decreases at each step of the optimization
process and converges to $\mu^*_{\rm app}$. The right derivative of
$\delta$ at $\mu^*_{\rm app}$ being always positive, 
the convergence rate is guaranteed to be at least superlinear. 

Of course, the most difficult part of the algorithm is the computation
of the distance $\delta(\mu)$ to the cone, and of the projection $A_\mu$
of $K_N-\mu$. To this end, we chose to minimize, for a given $\mu$, the
objective function  
\[
J_\mu(B) = \frac{1}{2}\left\|K_N-\mu-\sum_{\ell=1}^L(\cL_\ell)^*B_\ell\right\|^2,
\]
under the constraints $B_\ell \geq 0$ ($\ell=1...L$), according to the
definition \eqref{app_dual_cone} of the polar cone $(\cC_{\rm
  app})^*$. The above minimization is performed using a classical
limited-memory BFGS algorithm~\cite{opt}, keeping the last $m=3$ descent
directions. The positivity constraints were parametrized by $B_\ell =
(C_\ell)^2$ with $C_\ell$ symmetric, as suggested by Mazziotti in
\cite{mazziottiPRL,mazziotti04}.

Computing $\delta(\mu)$ with sufficient accuracy when $\mu$ is close to
$\mu^*_{\rm app}$ can be difficult because the  minimization of
$J_\mu(B)$ then is ill-conditioned. We therefore consider a ``truncated"
version of the Newton algorithm where $\mu$ is updated by a fraction $0
< a \le 1$ of the Newton step. We then use the linearity of $\delta$ for
values close 
to $\mu^*_{\rm app}$ to devise a stopping criterion limiting the number
of iterations. The algorithm is as follows: 
\begin{algo} 
Consider an initial value $\mu^0$ (for example the Hartree-Fock value
$\mu_{\rm HF}$), and $0 < a \leq1$. Compute the projection
$A_{\mu^0}$ of $K_N-\mu^0$ on $(\cC_{\rm app})^*$
and the distance $d^0=\delta(\mu^0)$, and
consider $\mu^1 = \mu^0 - \frac{\delta(\mu^0)}{\delta'(\mu^0)}$. For $n
\geq 1$, 
\begin{itemize}
\item Step 1. Compute the projection $A_{\mu^n}=\sum_{\ell=1}^L(\cL_\ell)^*\left[(C_\ell^n)^2\right]$
  of $K_N-\mu^n$ on $(\cC_{\rm app})^*$, the associated distance $d^n =
  \delta(\mu^n)=||K_N-\mu^n-A_{\mu^n}||$ and the derivative $\delta'(\mu^n)$; 
\item Step 2. Compute the interpolation slope $p^n = \frac{d^{n-1}-d^n}{\mu^{n-1}-\mu^n}$;
\item Step 3. If $p^n \leq (1 +\epsilon)\delta'(\mu^n)$, then the linear assumption is satisfied and the final value is extrapolated from the current position as $\mu^* = \mu^n - \frac{\delta(\mu^n)}{\delta'\mu^n)}$;
\item Step 4. Otherwise, set $\mu^{n+1} = \mu^n - a \frac{\delta(\mu^n)}{\delta'(\mu^n)}$ and start again from~(1) using as initial guess $C_\ell^{n+1}=C_\ell^n$ for any $\ell=1...L$.
\end{itemize}
\end{algo}

In practice, the above algorithm converges in a few iterations. The only time consuming step is the projection performed in Step 1. 
As described above, this projection is done iteratively by minimizing the objective function $J_\mu$ by a limited-memory BFGS algorithm.
The cost of one BFGS iteration scales as ${\rm O}(r^6)$. We did not observe a clear scaling of the number of BFGS iterations with respect to the basis set size. The memory requirements scale as ${\rm O}(r^4)$. Both computational time and memory requirements are comparable to those of~\cite{mazziotti04}.

\section{Numerical results}
\label{section_num}

We have tested the method on several molecules at equilibrium geometries using data from~\cite{geom_web}, for STO-6G and 6-31G basis sets. 
The results are reported in Table~\ref{tab:sto6g} and~\ref{tab:631G} respectively.  

\bigskip

{\bf Tables I and II}

\bigskip

The
reference Full CI (FCI) energies have been computed using
GAMESS~\cite{Gamess}. The correlation energies are recovered with a 
good accuracy. This is consistent with
previous results already obtained with different RDM
methods~\cite{NNEFNF01,mazziottiPRL,mazziotti04,ZBFOP,FBNOPYZ}. 

In general, we have observed that the function $\delta$ is almost linear
in quite large a right neighborhood of $\mu_{\rm app}^*$ (see
Figure \ref{courbe_N2}). One iteration of the Newton algorithm already
provides a very correct approximation of the exact RDM energy, even when
starting from the Hartree-Fock level. Usually, only 3 or 4 Newton
iterations are necessary to achieve convergence. Therefore,
the only limiting step of the method is the computation of the distance
$\delta(\mu)$ and of the projection $A_\mu$ of $K_N-\mu$ on the polar
cone. The method is very robust with respect to initial choices of the
energy $\mu^0$ and the matrices $C_k^0$. However, we have observed that
the computational time needed for finding the projection $A_\mu$ highly
depends on the quality of the initial guess. The choice of genuine
initial conditions is not obvious since we are manipulating abstract
objects (dual elements of 2-RDM). Some CPU times are reported in
Table~\ref{tab:CPU} for very crude initial conditions 
$C_k^0={\rm Id}$ and $\mu^0 \simeq 0.9 \mu_{\rm HF}$. 

\bigskip

{\bf Table III}

\bigskip

We would like to
underline that our projection
algorithm is far from being
optimal. There is clearly much room for improvement here. Let us also
mention that the curve $\mu \to \delta(\mu)$ can be easily
sampled using parallel computing (one value of $\mu$ per processor).

\begin{center}
\begin{figure}[h]
\includegraphics[angle=270,width=13cm]{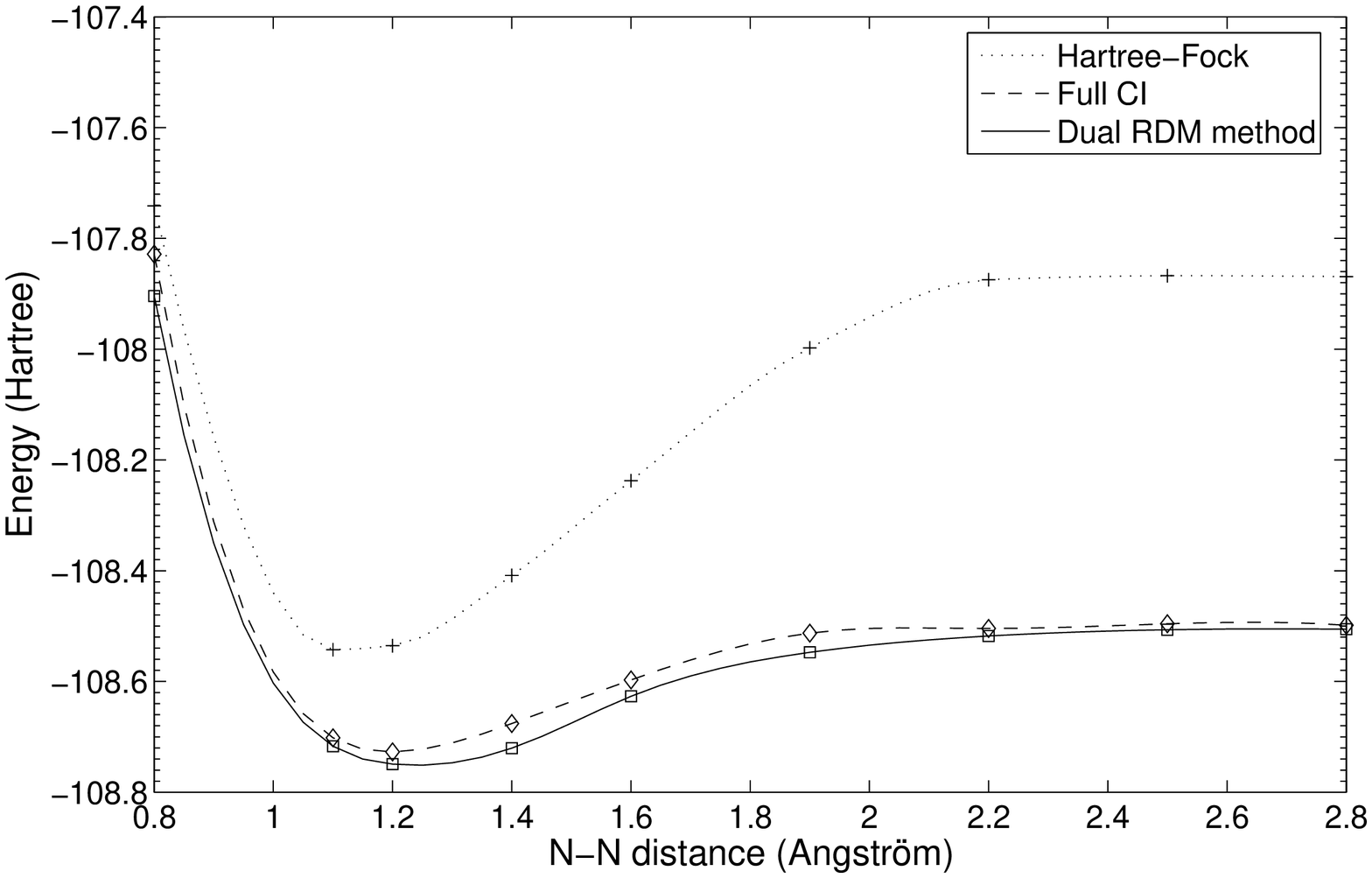}\hfill
\vspace{4cm}
\caption{ \label{dissociation} Cancès et al., Journal of Chemical Physics.}
\end{figure}
\end{center}

We also present in Figure~\ref{dissociation} dissociation curves for
N$_2$ in a STO-6G basis set. This example was already studied in several works~\cite{NEN02,JM04,GM05}. The
agreement of our results with the reference Full CI is excellent, and the
dissociation energy is therefore recovered with a very good accuracy. 

\bigskip

{\bf Figure II}

\bigskip

\section*{Acknowlegements}

We acknowledge the computer facilities of the University of Cergy. This work was supported by the ACI ``Molecular Simulation'' of the French Ministry of Research. Part of this work was done while G. Stoltz was attending the program ``Bridging Time and Length Scales" at IPAM (UCLA).
We also thank Dr. M. Nakata for interesting comments.


\newpage

\section*{Table and Figures captions}

\begin{itemize}
\item {\bf Figure I}. Left: Distance $\delta(\mu)$ of $K_N-\mu$ to the
  cone $(\cC_{\rm app})^*$ as a function of $\mu$ for N$_2$ in a STO-6G
  basis set.    
The tangent at the estimated value for $\mu_{\rm app}^*$ is also
  displayed (dotted line). Right: Zoom near the FCI reference value.
  The Hartree-Fock value is $\mu_{\rm HF} = -1.4435153$ while the
reference FCI value is $\mu_{\rm CI} = -1.4453909$.
\item {\bf Table I}. Correlation energies in a STO-6G basis set.
\item {\bf Table II}. Correlation energies in a 6-31G basis set.
\item {\bf Table III}. CPU time (s) in a STO-6G basis using very crude
  initial guesses ($C_l=I$).
\item {\bf Figure II}. Dissociation curve for N$_2$ in a STO-6G
  basis set.
\end{itemize}

\newpage

\newpage

\begin{table}
\caption{\label{tab:sto6g} Cancès et al., Journal of Chemical Physics.}
\vspace{2cm}
\begin{ruledtabular}
\begin{tabular}{cccc}
System & FCI energy & Correlation energy & Dual RDM energy ($\%$ of the correlation energy)\\
\hline
\hline
Be & -14.556086 & -0.0527274 & -14.556123 (100.07) \\
LiH & -7.972557 & -0.0190867 & -7.9727078 (100.79)  \\
BH & -25.058806 & -0.0569044 & -25.061771 (105.21) \\
Li$_2$ & -14.837571 & -0.0286889 & -14.839066 (105.21) \\
BeH$_2$ & -15.759498 & -0.0335151 & -15.761284 (105.33) \\
H$_2$O & -75.735839 & -0.0546392 & -75.738582 (105.02) \\
NH$_3$ & -56.0586005 & -0.0693410 & -56.074805 (123.37) \\
\hline
\end{tabular}
\end{ruledtabular}
\vspace{10cm}
\end{table}

\newpage

\begin{table}
\caption{\label{tab:631G} Cancès et al., Journal of Chemical Physics.}
\vspace{2cm}
\begin{ruledtabular}
\begin{tabular}{cccc}
System & FCI energy & Correlation energy & Dual RDM energy ($\%$ of the correlation energy)\\
\hline
\hline
Be &  -14.613545 & -0.0467812 & -14.613653 (100.23) \\
LiH & -7.995678 & -0.0185565 & -7.9959693 (101.57) \\
BH & -25.171730 & -0.0630461 & -25.176736 (107.94) \\
Li$_2$ & -14.893607 & -0.0277581 & -14.895389 (106.42) \\
BeH$_2$ & -15.798440 & -0.0402691 & -15.801066 (106.52) \\
H$_2$O & -76.120220 & -0.1401501 & -76.142125 (115.63) \\
NH$_3$ & -56.291315 & -0.1336141 & -56.318065 (120.02) \\
\hline
\end{tabular}
\end{ruledtabular}
\vspace{13cm}
\end{table}

\newpage

\begin{table}
\caption{\label{tab:CPU} Cancès et al., Journal of Chemical Physics.}
\vspace{2cm}
\begin{ruledtabular}
\begin{tabular}{cccc}
System & Spatial basis size $r$ & CPU time (s) & Newton iterations \\
\hline
\hline
Be & 5 & 25.7 & 2 \\
LiH & 6 & 240.9 & 3 \\
H$_2$O & 7 & 958.8 & 4 \\
BeH$_2$ & 7 & 1143.3 & 3\\
\hline
\end{tabular}
\end{ruledtabular}
\end{table}

\newpage

\end{document}